\documentclass[aps,prd,twocolumn,nofootinbib,superscriptaddress,10pt]{revtex4-1}
\usepackage{amsmath,amsthm,latexsym,amssymb,amsfonts}
\usepackage[parfill]{parskip}
\usepackage{accents}
\usepackage{graphicx,color}
\usepackage{hyperref,natbib}
\usepackage[utf8]{inputenc}
\usepackage[T1]{fontenc}
\usepackage{textcomp}

\numberwithin{equation}{section}

\begin{document}

\title{Deformed Weitzenb\"ock Connections and Teleparallel Gravity}

\author{Victor A. Penas}
\email{victor.penas@cab.cnea.gov.ar}
\affiliation{G. F\'isica CAB-CNEA and CONICET,\\ Centro At\'omico Bariloche, Av. Bustillo 9500, Bariloche, Argentina.} 

\begin{abstract}
 We study conditions on a generic connection written in terms of first-order derivatives of the vielbein in order to obtain (possible) equivalent theories to Einstein Gravity. We derive the equations of motion for these theories which are based on the new connections. We recover the Teleparallel Gravity equations of motion as a particular case. The analysis of this work might be useful to Double Field Theory to find other connections determined in terms of the physical fields. \\
 
 {\bf NOTE}: The full content of this article can be found in the published version ``Deformed Weitzenböck Connections and Double Field Theory,'' Fortsch.\ Phys.\  {\bf 67} (2019) no.3,  1800077 doi:10.1002/prop.201800077. [arXiv:1807.01144 [hep-th]]   
\end{abstract}

\maketitle

\section{Introduction}

Teleparallel Gravity (TG or TEGR)\footnote{Unless otherwise stated, in this work when we refer to TG we mean the standard teleparallel equivalent to general relativity.} is an alternative formulation to Einstein Gravity (EG)\cite{firstrefs,Hayashi:1979qx,Arcos:2005ec}. It is based on parallelizable manifolds equipped with the Weitzenb\"ock connection (the connection of the parallelization) rather than the Levi-Civita connection. One of the key aspects of the theory is that the Weitzenb\"ock connection furnishes a null Riemann curvature tensor, and since this connection is metric-compatible, the dynamics of the theory is based only on the torsion. The Weitzenb\"ock connection is of the form $W_{\mu \nu}{}^{\rho}= \partial_{\mu} e^{a}{}_{\nu}  e_{a}{}^{\rho}$, where $\mu$ and $a$ are curved and flat indices respectively, and $e_{a}{}^{\mu}$ is the vielbein. The vielbein is treated as the fundamental field of the theory and the metric is interpreted as a byproduct coming from it.      
The name of TG comes from the fact that there is a notion of absolute parallelism between vectors located at different spacetime points. Indeed, one way to see this is to notice that the vielbein field is covariantly constant with respect to the  Weitzenb\"ock connection in a coordinate basis. This means that we can compare the flat components of a vector field defined at different spacetime points just like we would do in a global basis of flat spacetime.

TG has been found to have applications in cosmology by studying theories called $f(T)$-Theories \cite{Ferraro:2006jd, Bengochea:2008gz, Linder:2010py, Bamba:2010wb, Li:2011wu, Cai:2015emx, Paliathanasis:2016vsw}, where $T$ stands for scalar combinations built out of the torsion. In general, the local Lorentz symmetry group of these theories is broken and only a remnant subgroup of it survives \cite{Li:2010cg, Sotiriou:2010mv, Ferraro:2014owa}. It is easy to see that spacetime tensors built out of the Weitzenb\"ock connection (for instance the torsion $T_{\mu\nu}{}^{\rho}$) will not always transform as scalars under local Lorentz transformations. In fact, the Weitzenb\"ock connection itself, in a coordinate basis, does not transform like a scalar under local Lorentz transformations but under global ones. Global transformations bring the immediate worry about extra degrees of freedom coming from the vielbein, since in $D=4$, six out of the 16 components of the vielbein are gauged away by local Lorentz transformations. However, since the TG action turns out to be equal to the Einstein-Hilbert action, up to a boundary term, the equations of motion of TG are equivalent to the Einstein equations and thus the theory enjoys local Lorentz symmetry. 

The TG idea has also been applied to Double Field Theory (DFT) \cite{Siegel:1993xq,Hull:2009mi,Aldazabal:2013sca}, which is a string-inspired field theory with manifest T-duality symmetry. 
In DFT, many geometric notions like diffeomorphisms, covariant derivatives (connections) and curvature tensors arise in a natural and generalized way. In fact, its geometric structure is closely related to generalized geometry \cite{Hitchin:2004ut}. However, in contrast to General Relativity, the DFT version of the Levi-Civita connection has not all of its components fully determined in terms of the physical fields, but only some projections of the connection are. This also implies that the DFT curvature tensor is not fully determined in terms of the physical fields \cite{Hohm:2011si}. In an attempt to determine the connection in terms of the physical fields, the TG version of DFT was considered \cite{Berman:2013uda}. The connection was set equal to the Wetizenb\"ock connection. The resulting theory, based on this connection, reproduces the DFT action (up to a boundary term) and the dynamics is based on the (generalized) torsion. Similarly as in TG, the DFT action possesses a local double Lorentz symmetry but the generalized torsion transforms only under global double Lorentz transformations.

Motivated by this global-local mechanism of TG, and inspired by the necessity of DFT to consider other connections determined in terms of the physical fields of the theory, we will try to obtain equivalent theories to EG defined by other connections rather than the Levi-Civita or the Weitzenb\"ock connection. These connections should be determined in terms of the vielbein and its derivative. Although these connections will transform under global Lorentz transformations, the resulting theories will possess local Lorentz symmetry. Our starting point will be the most general connection written in terms of first-order derivatives of the vielbein\footnote{We will not consider terms that could involve the epsilon tensor $\epsilon_{\mu\nu\rho\sigma}$.}: 
\begin{align}\nonumber
\Gamma_{\mu \nu}{}^{\rho}= W_{\mu\nu}{}^{\rho} - {a}_{1} \overset{(W)}{T}_{\nu \mu}\,^{\rho} + {b}_{1} \overset{(W)}{T}{}^{\rho}\,_{\mu \nu} + {b}_{2} \overset{(W)}{T}{}^{\rho}\,_{\nu \mu}\\\label{GammaConnectioncurvedsix}
- {c}_{1} g_{\mu \nu} \overset{(W)}{T}_{\sigma}\,^{\rho \sigma} - {d}_{1} {\delta}_{\mu}\,^{\rho} \overset{(W)}{T}_{\nu}\,^{\sigma}\,_{\sigma} - {d}_{2} {\delta}_{\nu}\,^{\rho} \overset{(W)}{T}_{\mu}\,^{\sigma}\,_{\sigma}.
\end{align}
Where $\overset{(W)}{T}_{\mu\nu}{}^{\rho}=2W_{[\mu\nu]}{}^{\rho}$ is the torsion of the Weitzenb\"ock connection\footnote{We are using the following conventions on indices: $(a b)=\frac{1}{2}\Big(a b + b a\Big)$ and $[a b]=\frac{1}{2}\Big(a b - b a\Big)$. In section (\ref{eqforCoeff}) we will denote quantities associated with the Weitzenb\"ock connection with a $(4)$ as superscript, e.g. $\overset{(W)}{T}_{\mu\nu}{}^{\rho}=\overset{(4)}{T}_{\mu\nu}{}^{\rho}$.} and $g_{\mu\nu}$ is the spacetime metric. This connection is parametrized by six real parameters and it does not have any symmetry requirement imposed on its indices so it will yield, generically, torsion and non-metricity. In a coordinate basis, this connection transforms in  a proper way under general coordinate transformations and is a scalar under \emph{global} Lorentz transformations. This means that for generic coefficients, the local Lorentz symmetry is broken (see, however, below equation (\ref{SpinConnectionTau})).  

The rest of the paper is organized as follows. In section (\ref{Notation}) we will introduce the notation used in the paper and show how to obtain different actions equivalent to the Einstein-Hilbert action for connections with generic features.  We will not talk about matter coupling so the theories obtained from the generic connections will be equivalent to EG in vacuum. 
In section (\ref{eqforCoeff}) we will find constraints on the parameters of (\ref{GammaConnectioncurvedsix}). The parameters that satisfy these constraints will give the desired connections that allow for actions equivalent, up to a boundary term, to the Einstein-Hilbert action and we will analyze particular values of these parameters. In section (\ref{eqofmotionsection}) we will obtain the equations of motion for these theories based on the new connections. In section (\ref{conclusions}) we summarize our work.

\section{Notation and Idea}\label{Notation}

We will work in $D=4$ spacetime dimensions, $\mu,\nu...$ are curved indices and $a,b...$ denote flat indices. Our fundamental field is the vielbein $e_{a}{}^{\mu}$ and we define the metric as a byproduct $g_{\mu \nu}=\eta_{a b} e^{a}{}_{\mu} e^{b}{}_{\nu}$. Here $\eta_{a b}$ is the constant Minkowski metric. 
The vielbein satisfies $e_{a}{}^{\mu} e^{b}{}_{\mu} = \delta_{a}{}^{b}$ and $e^{a}{}_{\mu} e_{a}{}^{\nu} = \delta_{\mu}{}^{\nu}$ where we raise and lower curved and flat indices with the metric $g_{\mu \nu}$ and $\eta_{a b}$ respectively. We define the flat derivative $D_{a}=e_{a}{}^{\mu}\partial_{\mu}$. It is useful to introduce the following quantities:
\begin{equation}
\Omega_{a b}{}^{c}=D_{a} e_{b}{}^{\rho} e^{c}{}_{\rho},~~~~\tau_{a b}{}^{c}=2\Omega_{[a b]}{}^{c}.
\end{equation}
Here $\tau_{a b}{}^{c}$ are the anholonomy coefficients. The Riemann curvature tensor is defined as follows:
\begin{equation}\label{RiemannCurved}
R_{\mu \nu \rho}{}^{\sigma}=\partial_{\mu} \Gamma_{\nu \rho}{}^{\sigma} - \partial_{\nu} \Gamma_{\mu \rho}{}^{\sigma} - \Gamma_{\mu \rho}{}^{\lambda} \Gamma_{\nu \lambda}{}^{\sigma} + \Gamma_{\nu \rho}{}^{\lambda} \Gamma_{\mu \lambda}{}^{\sigma}. 
\end{equation}
For generic affine connections the Riemann tensor is only antisymmetric in the first two indices. In a metric-affine spacetime the affine connection admits a general decomposition of the form:
\begin{equation}\label{eqDecompOfGamma}
\Gamma_{\mu \nu}{}^{\sigma}= \mbox{\scriptsize{$\left\{\begin{array}{@{}c@{}} \sigma \\ \mu ~\nu\end{array}\right\}$}} + K_{\mu \nu}{}^{\sigma} + L_{\mu \nu}{}^{\sigma},
\end{equation}
where the Christoffel symbols, contorsion tensor and the $L_{\mu\nu}{}^{\rho}$ tensor are given respectively by:
\begin{equation}
\mbox{\scriptsize{$\left\{\begin{array}{@{}c@{}} \sigma \\ \mu ~\nu\end{array}\right\}$}}=\frac{1}{2} g^{\tau \sigma}\left(\partial_{\mu} g_{\nu \tau} - \partial_{\tau} g_{\mu \nu} + \partial_{\nu} g_{\tau \mu}\right), 
\end{equation}
\begin{equation}
K_{\mu \nu}{}^{\sigma}=-\frac{1}{2}g^{\tau \sigma}\left(T_{\mu \tau}{}^{\rho} g_{\rho \nu} - T_{\tau \nu}{}^{\rho} g_{\rho \mu} - T_{\mu \nu}{}^{\rho} g_{\rho \tau}\right),
\end{equation}
\begin{equation}
L_{\mu \nu}{}^{\sigma}=-\frac{1}{2} g^{\tau \sigma}\left(\nabla_{\mu} g_{\nu \tau} - \nabla_{\tau} g_{\mu \nu} + \nabla_{\nu} g_{\tau \mu}\right). 
\end{equation}
Here, the torsion is $T_{\mu\nu}{}^{\sigma}=2\Gamma_{[\mu\nu]}{}^{\sigma}$ and the covariant derivative is defined as $\nabla_{\mu}V_{\nu}=\partial_{\mu} V_{\nu} - \Gamma_{\mu \nu}{}^{\rho} V_{\rho}$. The combination $Q_{\mu\nu\rho}=\nabla_{\mu}g_{\nu\rho}$ is known as the non-metricity tensor. When we plug the above decomposition in (\ref{RiemannCurved}) we obtain:
\begin{align}\nonumber
R_{\mu\nu\rho}{}^{\sigma}(\Gamma)&=R_{\mu\nu\rho}{}^{\sigma}(\{\}) + \overset{(\{\})}{T}_{\mu \nu}{}^{\lambda}C_{\lambda\rho}{}^{\sigma} \\\label{eqDecompOfRiemann}
&~~~~+ 2\overset{(\{\})}{\nabla}_{[\mu} C_{\nu]\rho}{}^{\sigma} + 2C_{[\mu|\lambda|}{}^{\sigma}C_{\nu]\rho}{}^{\lambda}\, ,
\end{align}
where $C_{\mu\nu}{}^{\sigma}=K_{\mu\nu}{}^{\sigma} + L_{\mu\nu}{}^{\sigma}$ and $\overset{(\{\})}{\nabla}_{\mu}$ is referred to the Levi-Civita connection. The Ricci tensor and Ricci scalar are defined as
\begin{equation}
R_{\mu \nu}=R_{\mu \rho \nu}{}^{\rho}, ~~~~R=g^{\mu \nu}R_{\mu \rho \nu}{}^{\rho}.
\end{equation}
The decomposition (\ref{eqDecompOfRiemann}) on the Ricci scalar yields:
\begin{align}\nonumber
R(\Gamma)&=R(\{\}) + \overset{(\{\})}{\nabla}_{\mu}C_{\sigma}{}^{\mu\sigma} - \overset{(\{\})}{\nabla}_{\sigma}C_{\mu}{}^{\mu\sigma}\\\label{Rscalardecomposed}
&~~~~~~~~~~~~~~~~ + C_{\mu\lambda}{}^{\sigma} C_{\sigma}{}^{\mu\lambda} - C_{\sigma\lambda}{}^{\sigma}C_{\mu}{}^{\mu\lambda}.
\end{align} 
The Levi-Civita connection is torsionless and metric-compatible, which means that $\overset{(\{\})}{T}_{\mu\nu}{}^{\sigma}=0$ and the metric can pass through the covariant derivative.

The relation between the components of the affine connection written in a holonomic basis and an anholonomic one is of the form:
\begin{equation}\label{affineandgaugeconn}
\Gamma_{\mu \nu}{}^{\rho}=W_{\mu \nu}{}^{\rho} + e ^{b}{}_{\nu} e_{c}{}^{\rho} w_{\mu b}{}^{c}.
\end{equation} 
For instance, if $\Gamma_{\mu\nu}{}^{\rho}$ is the Levi-Civita connection then $w_{\mu a}{}^{b}$ is the usual spin connection (\ref{SpinConnLevCivTau}). For simplicity, we will refer to $w_{\mu a}{}^{b}$ as a gauge connection. By inserting (\ref{GammaConnectioncurvedsix}) in (\ref{affineandgaugeconn}) and using $W_{\mu \nu}{}^{\rho}=-e^{a}{}_{\mu} e^{b}{}_{\nu} e_{c}{}^{\rho} \Omega_{a b}{}^{c}$ the gauge connection gets the following form: 
\begin{align}\nonumber
w_{a b}{}^{c} = & \, {a}_{1} {\tau}_{b a}\,^{c} - {b}_{1} {\tau}^{c}\,_{a b} - {b}_{2} {\tau}^{c}\,_{b a} + {c}_{1} {\eta}_{a b} {\tau}_{d}\,^{c d}\\\label{SpinConnectionTau}
& + {d}_{1} {\delta}_{a}\,^{c} {\tau}_{b}\,^{d}\,_{d} + {d}_{2} {\delta}_{b}\,^{c} {\tau}_{a}\,^{d}\,_{d}.
\end{align}
The above gauge connection transforms as a scalar under diffeomorphisms. However, for generic coefficients, it does not transform like a gauge connection under local gauge transformations, but only global ones. In fact, if we force this connection to transform under local Lorentz transformations we see that the only possibility is with $a_{1}=b_{1}=b_{2}=-1/2$,  $c_{1}=d_{1}=d_{2}=0$. This is precisely the coefficients for the Levi-Civita spin connection:
\begin{align}\label{SpinConnLevCivTau}
\overset{(\{\})}{w}_{a b}{}^{c}= \frac{1}{2}\, {\tau}_{a b}\,^{c} + \frac{1}{2}\, {\tau}^{c}\,_{a b} + \frac{1}{2}\, {\tau}^{c}\,_{b a}.
\end{align}
In planar indices, the Riemman tensor (\ref{RiemannCurved}) takes the form:
\begin{align}\nonumber
R_{a b c}{}^{d} = & {D}_{a}{{w}_{b c}\,^{d}}\,  - {D}_{b}{{w}_{a c}\,^{d}}\,  - {w}_{a c}\,^{e} {w}_{b e}\,^{d} \\\label{RiemannFlat}
&+ {w}_{b c}\,^{e} {w}_{a e}\,^{d} - {\tau}_{a b}\,^{e} {w}_{e c}\,^{d}.
\end{align}
Similarly as before, the Ricci tensor and the Ricci scalar are of the form $R_{a b}=R_{a c b}{}^{c}$ and  $R=\eta^{a b}R_{a b}$ respectively.

The idea is to use the connection (\ref{SpinConnectionTau}) (or equivalently (\ref{GammaConnectioncurvedsix})) and find conditions on the coefficients ($a_{1}$, $b_{1}$, $b_{2}$, $c_{1}$, $d_{1}$, $d_{2}$) such that the Ricci scalar for this connection vanishes. In this way, we obtain an equality between the usual Einstein-Hilbert action and an action with $C$-terms (up to a boundary term). Indeed, consider the Einstein-Hilbert action with the scalar curvature (\ref{Rscalardecomposed}):
\begin{align}\nonumber
&\int dx^4\sqrt{-g}R(\Gamma)=  \int dx^4 \sqrt{-g}\Big(R(\{\}) \\\label{FundamentalEquationone}
& ~~~~~~~~~~~~~~~~~~~~~+ C_{\mu\lambda}{}^{\sigma}C_{\sigma}{}^{\mu\lambda} - C_{\sigma\lambda}{}^{\sigma}C_{\mu}{}^{\mu\lambda}\Big).
\end{align}
We have dropped out the covariant derivative terms since they form a total derivative.
By setting $R(\Gamma)=0$ in (\ref{FundamentalEquationone}) we simply obtain:
\begin{align}\nonumber
&\int dx^{4} \sqrt{-g}R(\{\}) = \\\label{FundamentalEquationtwo}
&~~~~~=- \int dx^{4} \sqrt{-g}\left(C_{\mu\lambda}{}^{\sigma}C_{\sigma}{}^{\mu\lambda} - C_{\sigma\lambda}{}^{\sigma}C_{\mu}{}^{\mu\lambda}\right).
\end{align}

We stress again that the above equality should be understood up to a boundary term. In the second line of (\ref{FundamentalEquationtwo}), the tensor $C_{\mu\nu}{}^{\rho}$ depends on the general connection $\Gamma$ (\ref{GammaConnectioncurvedsix}). This means that  connections that furnish a null Ricci scalar can be used to yield an action equivalent to the Einstein-Hilbert action by using its torsion and/or non-metricity. This is indeed the case for the usual TG, where the right hand side of (\ref{FundamentalEquationtwo}) reproduces exactly the teleparallel action with $C_{\mu\nu}{}^{\rho}(W)=K_{\mu\nu}{}^{\rho}(W)$ (see section \ref{subunoEOMsec}).

The reader might be worried about the fact that we are considering global Lorentz transformations. But all of the connections we will obtain reproduce theories which possess local gauge transformations as established by the action (\ref{FundamentalEquationtwo}), reducing in this way the number of degrees of freedom of the vielbein. This is the same mechanism as in teleparallel gravity\footnote{At least for the usual teleparallel gravity which is equivalent to Einstein gravity (TEGR). As mentioned in the introduction, the Lorentz transformations in $f(T)$-gravities are more subtle.} where the affine connection is chosen to be the Weitzenb\"ock connection $W_{\mu\nu}{}^{\rho}$ which is invariant only under global Lorentz transformations but the resulting theory turns out to have local Lorentz symmetry.   

For simplicity we will mostly work with flat indices.  
The usual Einstein-Hilbert action with Levi-Civita connection after a partial integration is of the form: 
\begin{align}\nonumber
\int dx^{4} \sqrt{-g} R(\{\})= & \int dx^{4} \sqrt{-g} \Big(-\tau_{a c}{}^{c} \tau^{a b}{}_b\, \\
&+ \frac{1}{2}\, {\tau}^{a}\,_{b}\,^{c} {\tau}_{a c}\,^{b} + \frac{1}{4}\, {\tau}^{a b}\,_{c} {\tau}_{a b}\,^{c} \Big),
\end{align}
where we have used (\ref{RiemannFlat}) and (\ref{SpinConnLevCivTau}). Equation (\ref{FundamentalEquationtwo}) then gets the form:

\begin{align}\nonumber
\int dx^{4} \sqrt{-g} \Big(&-\tau_{a c}{}^{c} \tau^{a b}{}_b\,  + \frac{1}{2}\, {\tau}^{a}\,_{b}\,^{c} {\tau}_{a c}\,^{b}  + \frac{1}{4}\, {\tau}^{a b}\,_{c} {\tau}_{a b}\,^{c} \Big)\\ \label{FundamentalEquationthree}
&~= - \int dx^{4} \sqrt{-g}\Big(C_{a b}{}^{c}C_{c}{}^{a b} - C_{c b}{}^{c}C_{a}{}^{a b}\Big).
\end{align}

\section{Equations for coefficients}\label{eqforCoeff}

As mentioned before we would like to find conditions on the coefficients of the general connection (\ref{SpinConnectionTau}) such that the Ricci scalar of this connection vanishes. We plug (\ref{SpinConnectionTau}) in the Riemann tensor (\ref{RiemannFlat}) and contract indices to obtain the Ricci scalar. The result is the following:
\begin{align}\nonumber
R(w)=& {D}^{a}{{\tau}_{a}\,^{b}\,_{b}}\,  \Big({a}_{1} + 2\, {b}_{2} + 3\, {c}_{1} + 3\, {d}_{1} + {b}_{1}\Big) \\\nonumber
&+ {\tau}^{a b c} {\tau}_{a b c} \Big({a}_{1} {b}_{1} - {b}_{1} {b}_{2} - {a}_{1} {b}_{2} - {b}_{2}\Big)\\\nonumber
& + {\tau}^{a b c} {\tau}_{a c b} \Big( - {a}_{1} {a}_{1} + {a}_{1} {b}_{2} - {a}_{1} {b}_{1} - {b}_{1} {b}_{1} \\\nonumber
&~~~~~~~~~~~~~~~~~ + {b}_{1} {b}_{2} - {b}_{2} {b}_{2} - {a}_{1} - {b}_{1}\Big)\\\nonumber
&  + {\tau}_{b}\,^{c}\,_{c} {\tau}^{a b}\,_{a} \Big( - {a}_{1} {b}_{1} - {b}_{1} {b}_{2} + 2\, {b}_{1} {c}_{1} - 4\, {b}_{1} {d}_{1} \\\nonumber
& ~~~~~~~~~~~~~~~~~~ - {a}_{1} {b}_{2} - {b}_{2} {b}_{2} - 2\, {b}_{2} {c}_{1} - 2\, {b}_{2} {d}_{1}\\\nonumber
&~~~~~~~~~~~~~~~~~~ - 4\, {a}_{1} {c}_{1} + 3\, {c}_{1} {c}_{1} - 12\, {c}_{1} {d}_{1}\\\label{RicciScalarSpinConnTau}
&~~~~~~~~~~~~~~~~~~ + 2\, {a}_{1} {d}_{1} + 3\, {d}_{1} {d}_{1} + {c}_{1} + {d}_{1}\Big);
\end{align}

All of the terms are independent of each other so in order to obtain a null Ricci scalar we must set each parentheses to zero. 
This defines a set of four quadratic equations for five variables ($a_{1}$, $b_{1}$, $b_{2}$, $c_{1}$, $d_{1}$). Note that the coefficient $d_{2}$ has dropped out from (\ref{RicciScalarSpinConnTau}) but does not drop out from the Riemann tensor nor the Ricci tensor. The reason is the following. It is known that a Riemann tensor with a generic connection is invariant under:
\begin{equation}\label{gaugetransconncurvature}
\Gamma_{\mu\nu}{}^{\rho}\rightarrow \Gamma_{\mu\nu}{}^{\rho} + \delta_{\nu}{}^{\rho}\partial_{\mu}\phi,
\end{equation}
where $\phi$ is a scalar. However, the Ricci scalar built from this generic Riemann tensor is invariant under a relaxed version of the above equation: 
\begin{equation}\label{gaugetransfRicciScalar}
\Gamma_{\mu\nu}{}^{\rho}\rightarrow \Gamma_{\mu\nu}{}^{\rho} +A_{\mu}\delta_{\nu}{}^{\rho}.
\end{equation}
This is due to a contraction between the metric and $\partial_{[\mu}A_{\nu]}$. We can see that the term with coefficient $d_{2}$ in  (\ref{GammaConnectioncurvedsix}) has exactly the same form as in (\ref{gaugetransfRicciScalar}). We will come back to this point later.
 
We might have set to zero the Riemman tensor or the Ricci tensor in order to obtain equations for the coefficients. Nevertheless, we analyze the vanishing of the Ricci scalar because the solutions for this one includes the cases for a null Riemman and Ricci-flat tensors. Therefore, teleparallel gravity must show up as a particular solution to our set of four quadratic equations derived from (\ref{RicciScalarSpinConnTau}).\\ 

Our set of four equations have an infinite amount of solutions yielding both torsion and non-metricity (and, in general, a non-vanishing curvature). To simplify the analysis we consider particular cases in the following subsections.

\subsection{Metric-Compatible case}\label{eqforCoeffMCcase}

In this subsection we restrict to the case of metric-compatible connections, i.e.:
\begin{equation}
D_a \eta_{a b} - w_{a b}{}^{e}\eta_{e c} - w_{a c}{}^{e}\eta_{b e} = 0 ~~\Rightarrow ~~ w_{a (b c)} = 0. 
\end{equation}
The above condition implies 
\begin{equation}\label{condmetriccompat}
a_{1}=b_{1},~~c_{1}=d_{1},~~d_{2}=0,
\end{equation}
The set of four equations derived from (\ref{RicciScalarSpinConnTau}) together with conditions (\ref{condmetriccompat}) yield only four solutions which will be referred to as cases 1 to 4. These are respectively:

\begin{equation}\label{QzeroSolOne}
a_{1}=-1,~~ b_{2}=-1, ~~c_{1}= 2/3,
\end{equation}
\begin{equation}\label{QzeroSolTwo}
a_{1}=-2/3,~~ b_{2}=-4/3,~~ c_{1}= 2/3,
\end{equation}
\begin{equation}\label{QzeroSolThree}
a_{1}=-1/3,~~ b_{2}=1/3,~~ c_{1}=0,
\end{equation}
\begin{equation}\label{QzeroSolFour}
a_{1}=0,~~ b_{2}=0,~~ c_{1}=0,
\end{equation}
Replacing cases 1 to 4 in the general connection (\ref{GammaConnectioncurvedsix}) we obtain the following connections, respectively:  
\begin{align}\nonumber
\overset{(1)}{\Gamma}_{\mu \nu}{}^{\rho}= W_{\mu\nu}{}^{\rho} &+ \overset{(W)}{T}_{\nu \mu}\,^{\rho} - \overset{(W)}{T}{}^{\rho}\,_{\mu \nu} - \overset{(W)}{T}{}^{\rho}\,_{\nu \mu}\\\label{QzeroSolOneConn}
&- \frac{2}{3} g_{\mu \nu} \overset{(W)}{T}_{\sigma}\,^{\rho \sigma} - \frac{2}{3} {\delta}_{\mu}\,^{\rho} \overset{(W)}{T}_{\nu}\,^{\sigma}\,_{\sigma}.
\end{align}
\begin{align}\nonumber
\overset{(2)}{\Gamma}_{\mu \nu}{}^{\rho}= W_{\mu\nu}{}^{\rho} &+\frac{2}{3} \overset{(W)}{T}_{\nu \mu}\,^{\rho} -\frac{2}{3} \overset{(W)}{T}{}^{\rho}\,_{\mu \nu} -\frac{4}{3} \overset{(W)}{T}{}^{\rho}\,_{\nu \mu}\\\label{QzeroSolTwoConn}
&- \frac{2}{3} g_{\mu \nu} \overset{(W)}{T}_{\sigma}\,^{\rho \sigma} - \frac{2}{3} {\delta}_{\mu}\,^{\rho} \overset{(W)}{T}_{\nu}\,^{\sigma}\,_{\sigma}.
\end{align}
\begin{align}\nonumber
\overset{(3)}{\Gamma}_{\mu \nu}{}^{\rho}= W_{\mu\nu}{}^{\rho} &+\frac{1}{3} \overset{(W)}{T}_{\nu \mu}\,^{\rho} -\frac{1}{3} \overset{(W)}{T}{}^{\rho}\,_{\mu \nu} + \frac{1}{3} \overset{(W)}{T}{}^{\rho}\,_{\nu \mu}.\\\label{QzeroSolThreeConn}
\end{align}
\begin{align}\label{QzeroSolFourConn}
\overset{(4)}{\Gamma}_{\mu \nu}{}^{\rho}= W_{\mu\nu}{}^{\rho}.
\end{align}
Equivalently, we use (\ref{SpinConnectionTau}) with the above coefficients.  As expected, one of the solutions (eq. (\ref{QzeroSolFourConn})) is the Weitzeb\"ock connection.

The above connections (\ref{QzeroSolOneConn})-(\ref{QzeroSolFourConn}) are metric compatible (the non-metricity tensor vanishes) and present non-vanishing torsion. Thus, only the contorsion tensor $K_{a b}{}^{c}$ will contribute to (\ref{FundamentalEquationthree}).
The contorsion tensor for the metric-compatible case is equal to: 
\begin{equation}\label{defcontorsionMCcase}
K_{a b}{}^c=C_{a b}{}^c=w_{a b}{}^c - \overset{(\{\})}{w}_{a b}{}^c.
\end{equation}
It is straightforward to check that (\ref{FundamentalEquationthree}) is satisfied for the four cases when plugging (\ref{defcontorsionMCcase}) in (\ref{FundamentalEquationthree}). That is:
\begin{align}\nonumber
& \overset{(i)}{K}_{a b}\,^{e} \overset{(i)}{K}_{e c}\,^{b} {s}^{c a} - \overset{(i)}{K}_{a b}\,^{a} \overset{(i)}{K}_{e c}\,^{b} {s}^{c e}=\\  &~~~~~~=- \frac{1}{2}\, {\tau}^{a b c} {\tau}_{a c b} - \frac{1}{4}\, {\tau}^{a b c} {\tau}_{a b c} + {\tau}^{b a}\,_{a} {\tau}_{b}\,^{c}\,_{c},
\end{align}
with $i=1,\cdots,4$ labeling cases 1 to 4.
We want to remark that the above connections not only yield non-vanishing torsion but also non-vanishing Riemman curvature and Ricci tensors. The only exception being case 4, i.e. the Weitzenb\"ock case $\overset{(4)}{\Gamma}_{\mu\nu}{}^{\rho}=W_{\mu\nu}{}^{\rho}$ (only has non-vanishing torsion).
\subsection{Non-metricity case}\label{eqforCoeffNMcase}
We begin studying the simplest case of torsionless non-metric connection. The condition is:
\begin{equation}
T_{\mu \nu}{}^\sigma = 0 \Rightarrow \tau_{a b}{}^c = w_{a b}{}^c - w_{b a}{}^c,
\end{equation}
which implies that the coefficients must satisfy
\begin{equation}\label{nonmetrtorsionlesscase}
\begin{split}
1+2a_1=0\\
b_1-b_2=0\\
d_2-d_1=0.
\end{split}
\end{equation}
However, equations (\ref{nonmetrtorsionlesscase}) together with the set of four quadratic equations derived from (\ref{RicciScalarSpinConnTau}) yield no solution. 
There have been other attempts for constructing equivalent theories to EG based purely on non-metricity. For instance in \cite{Nester:1998mp}  a vanishing $\Gamma_{\mu\nu}{}^{\rho}$ was considered, allowing the theory to be described only in terms of the non-metricity tensor $Q_{\mu\nu\rho}=\partial_{\mu}g_{\nu\rho}$. This is equivalent as choosing $w_{a b}{}^{c}=e_{a}{}^{\mu}e_{b}{}^{\nu}e^{c}{}_{\rho}W_{\mu\nu}{}^{\rho}$. We believe that if we had allowed an extra arbitrary constant factor in front of $W_{\mu\nu}{}^{\rho}$ in (\ref{GammaConnectioncurvedsix}), then our procedure would have also yielded this gauge connection as a particular result. However, the transformation properties of this connection  under diffeomorphisms would have been compromised.\\

As said before there are infinite solutions to our set of four equations presenting, in a generic way, non-metricity and torsion. Just to analyze a particular case we study the condition for a Weyl's space: 
\begin{equation}
\nabla_{\mu} g_{\nu\rho} = - 2A_{\mu}g_{\nu\rho} ~~\Rightarrow ~~w_{a (b c)} = A_{a} \eta_{b c}.  
\end{equation}  
Where $A_{\mu}$ is a one-form field (Weyl's vector field). Since we want a connection only in terms of derivatives of the vielbein the only possibility is $A_{a}=\alpha \tau_{a}{}^{d}{}_{d}$ for some real constant $\alpha$. This implies the conditions
 \begin{equation}\label{weylconditioncoeff}
 	a_{1}=b_{1}, ~c_{1}=d_{1}, ~\alpha=d_{2}. 
 \end{equation}
 These  conditions on the coefficients are the same as the previous case (\ref{condmetriccompat}) except we have a non-trivial condition on $d_{2}$. Since $d_{2}$ does not show up in (\ref{RicciScalarSpinConnTau}), the solutions to the quadratic equations plus (\ref{weylconditioncoeff}) are the same as the previous case but with an extra term, i.e.:
\begin{equation}\label{gammaweylspace}
\overset{(i)}{\widetilde{\Gamma}}_{\mu\nu}{}^{\rho}=\overset{(i)}{\Gamma}_{\mu\nu}{}^{\rho} -\alpha \overset{(W)}{T}_{\mu}{}^{\sigma}{}_{\sigma}\delta_{\nu}{}^{\rho},
\end{equation}

with $i=1,\cdots,4$. Although these connections are different from (\ref{QzeroSolOneConn})-(\ref{QzeroSolFourConn}) due to the last term in (\ref{gammaweylspace}), we must notice that this extra term is just of the form (\ref{gaugetransfRicciScalar}) which leaves the Ricci scalar invariant. More explicitly: $A_{\mu}\delta_{\nu}{}^{\rho}=-\alpha \overset{(W)}{T}_{\mu}{}^{\sigma}{}_{\sigma}\delta_{\nu}{}^{\rho}$.

\section{Gauge redundancy and deformed Weitzenb\"ock connections}

One might be worry about the fact that the different connections we are obtaining are, in some sense, gauge artifacts of the Weitzenb\"ock connection. First of all, we would like to adopt the point of view that a gauge transformation in this context should be understood as a local transformation performed on the vielbein (since the vielbein is our physical field). We want to stress again that only when taking global Lorentz transformations, the connection $\Gamma_{\mu\nu}{}^{\rho}$ and tensors built out of it are well defined, in the sense that they transform as a scalar under Lorentz transformations. And only when considering actions or equations of motion the local Lorentz symmetry is restored back. This is completely analogous to the usual teleparallel gravity case. Thus, if we consider global Lorentz transformations it is clearly not possible to gauge-transform the different connections into the Weitzenb\"ock connection. Moreover, if we allowed for local  Lorentz transformations it is not possible neither. Take for instance the curvature tensor (\ref{RiemannFlat}) with the general connection (\ref{SpinConnectionTau}):

\begin{align}\nonumber
R_{a b c}{}^{d}(w)= &D_{a} \Big({a}_{1} {\tau}_{c b}\,^{d} - {b}_{1} {\tau}^{d}\,_{b c} - {b}_{2} {\tau}^{d}\,_{c b} + {c}_{1} {\eta}_{b c} {\tau}_{e}\,^{d e} \\\nonumber
&+ {d}_{1} {\delta}_{b}\,^{d} {\tau}_{c}\,^{e}\,_{e} + {d}_{2} {\delta}_{c}\,^{d} {\tau}_{b}\,^{e}\,_{e}\Big) \\\nonumber
&-D_{b} \Big({a}_{1} {\tau}_{c a}\,^{d} - {b}_{1} {\tau}^{d}\,_{a c} - {b}_{2} {\tau}^{d}\,_{c a}\\\nonumber
& + {c}_{1} {\eta}_{a c} {\tau}_{e}\,^{d e} + {d}_{1} {\delta}_{a}\,^{d} {\tau}_{c}\,^{e}\,_{e} + {d}_{2} {\delta}_{c}\,^{d} {\tau}_{a}\,^{e}\,_{e}\Big)\\\label{fullRiemmangaugeconn}
 &+ \cdots,
\end{align}   
where the dots represent quadratic terms. From the above derivative terms we see that the only possibility for the curvature tensor to vanish is that all of the coefficients must be set equal to zero, implying that the Weitzenb\"ock connection is the only connection written in terms of derivatives of the vielbein that makes the curvature tensor to vanish. So, if the connections we have obtained so far were gauge-transformed (under \emph{local} Lorentz transformations) of the Weitzenb\"ock connection, it would imply a vanishing curvature tensor for these connections. As we have stated in the previous examples, this is not the case. More generally, different set of coefficients live in different Lorentz gauge orbits.   

On the other hand, we have mentioned about transformations that leave the Riemann tensor or Ricci scalar invariant, (\ref{gaugetransconncurvature}) and (\ref{gaugetransfRicciScalar}) respectively. From a geometric point of view, the connections are meant to be defined up to a ``gauge transformation''of the form (\ref{gaugetransconncurvature}) as long as the new connection enjoys the same index symmetry properties as the old one. For instance, the Levi-Civita connection is symmetric in its two lower indices and if we apply the transformation (\ref{gaugetransconncurvature}) the symmetry in the two indices will be broken resulting in a new connection with torsion and non-metricity. The connections obtained in this work do not have a particular definite symmetry on their indices, and thus, transformations of the form (\ref{gaugetransconncurvature}) might be considered. However, none of our connections are related in this sense, since there is no combination of vielbein that produces a non-trivial scalar as in the last term of (\ref{gaugetransconncurvature}). The only transformation relevant for us is (\ref{gaugetransfRicciScalar}). In this case there are connections that are related as we have explicitly shown in (\ref{gammaweylspace}). Despite of this, the geometric properties defined by $\overset{(i)}{\widetilde{\Gamma}}$ are different from $\overset{(i)}{\Gamma}$: Clearly, $\overset{(i)}{\widetilde{\Gamma}}$ yields non-trivial non-metricity while $\overset{(i)}{\Gamma}$ does not. Also, the form of the Riemman tensor for these connections are different since the $d_{2}$ term does not decouple from the Riemman tensor (see (\ref{fullRiemmangaugeconn})).

Having clarified the gauge redundancy issue, the interpretation we are giving to the new connections is that they are deformed versions of the Weitzenb\"ock connection \cite{Hehl:1994ue}. Given two connections $\Gamma_{1}$ and $\Gamma_{2}$ we can always relate them by a tensor. Take $\Gamma_{1}=W + w_{1}$ and $\Gamma_{2}=W + w_{2}$ as in the decomposition (\ref{affineandgaugeconn}). Since $w$ is a tensor under diffeomorphisms, we obtain  $\Gamma_{1}=\Gamma_{2} + (w_{1} - w_{2})$. Thus, $\Gamma_{1}$ and $\Gamma_{2}$ are related by a tensor and we say $\Gamma_{1}$ is a deformed version of $\Gamma_{2}$. In this work, what we did is to find those $w$ parametrized by (\ref{SpinConnectionTau}) such that $\Gamma_{1}=W + w$ is a deformation of $\Gamma_{2}=W$ and reproduce the same action (or equations of motion) for the vielbein.

\section{Equations of Motion}\label{eqofmotionsection}

In this section we will obtain the equations of motion for the vielbein in terms of quantities associated to the connections we have obtained in the previous sections. 

As a warm-up, we first obtain the equations of motion in the teleparallel case (i.e. $\overset{(4)}{w}_{a b}{}^{c}=0$) as it is usually done in the literature. However, we will apply a more general procedure in (\ref{subsecBGenericCaseEOM}) that will include the Weitzenb\"ock case as a particular case.

\subsection{Weitzenb\"ock case}\label{subunoEOMsec}

We are interested in the right hand side of (\ref{FundamentalEquationtwo}) using planar indices and the Weitzenb\"ock connection $\overset{(4)}{w}_{a b}{}^{c}=0$. For metric-compatible connections the non-metricity tensor vanishes, thus the only contribution to the action comes from the contorsion tensor, i.e. $C_{a b}{}^{c}=K_{a b}{}^{c}$. Expanding in terms of the torsion we get:

\begin{align}\nonumber
&\overset{(4)}{K}_{a b}{}^{c}\overset{(4)}{K}_{c}{}^{a b} - \overset{(4)}{K}_{c b}{}^{c}\overset{(4)}{K}_{a}{}^{a b}=\\
&~~=-\frac{1}{4}\overset{(4)}{T}_{a b c} \overset{(4)}{T}{}^{a b c} -\frac{1}{2}\overset{(4)}{T}_{a b c} \overset{(4)}{T}{}^{a c b} + \overset{(4)}{T}_{a b}{}^{b} \overset{(4)}{T}{}^{a c}{}_{c}.
\end{align}
Where we recall that $\overset{(4)}{T}_{a b}{}^{c}=-\tau_{a b}{}^{c}$. Since the Riemann tensor associated to the Weitzenb\"ock connection vanishes, we expect to find equations of motion in terms of only $T_{a b}{}^{c}$. The right hand side of (\ref{FundamentalEquationtwo}) thus gives:
\begin{align}\nonumber
& \int dx^{4} e\Big(-\frac{1}{4}\overset{(4)}{T}_{a b c} \overset{(4)}{T}{}^{a b c} -\frac{1}{2}\overset{(4)}{T}_{a b c} \overset{(4)}{T}{}^{a c b} + \overset{(4)}{T}_{a b}{}^{b} \overset{(4)}{T}{}^{a c}{}_{c}\Big)=\\
&~~~= \int dx^{4} e\Big(-\frac{1}{2}\overset{(4)}{\widehat{T}}{}^{a b c} \overset{(4)}{T}_{a b c}\Big).
\end{align}
Here $e=\sqrt{-g}$ and in the last line we have introduced the tensor $\widehat{T}_{a b c}$ known as the superpotential
\begin{align}\nonumber
\hat{T}_{a b c}&=K_{c b a} + \eta_{a c} T_{b d}{}^{d} - \eta_{b c} T_{a d}{}^{d}\label{superpotential}\\
& = \frac{1}{2}\Big(T_{a b c} + 2 T_{[a |c| b]} + \eta_{a c} T_{b d}{}^{d} - \eta_{b c} T_{a d}{}^{d} \Big).
\end{align}
The superpotential satisfies $\widehat{T}_{a b c} = 2 \widehat{T}_{[a b] c}$. Due to this property the variation of the lagrangian is easier to perform:
\begin{align}\nonumber
\delta_{e}\Big(-\frac{1}{2} e \overset{(4)}{\widehat{T}}{}^{a b c} \overset{(4)}{T}_{a b c} \Big) =&\frac{1}{2}e e_{d\mu}\delta e^{d\mu} \overset{(4)}{\widehat{T}}{}^{a b c} \overset{(4)}{T}_{a b c} \\
&~~~~ - e \overset{(4)}{\widehat{T}}{}^{a b c} \delta_e \overset{(4)}{T}_{a b c},
\end{align}
where
\begin{align}\nonumber
- e \overset{(4)}{\widehat{T}}{}^{a b c} \delta_e \overset{(4)}{T}_{a b c} \, = & \,  e \overset{(4)}{\widehat{T}}{}^{a b c} \Big(2\delta e_{a}{}^{\mu}e_{d \mu} \Omega^{d}{}_{b c} + \tau_{a b}{}^{d} e_{d}{}^{\mu} \delta e_{c \mu}\\ \label{stvofaction}
& ~~~~~~~~~~~~~~~~ + 2 D_{a}\delta e_{b}{}^{\mu} e_{c \mu} \Big)
\end{align}
Since the last term of (\ref{stvofaction}) is antisymmetrized on indices $(a,b)$ we can use:
\begin{equation}\label{DaCalDa}
2D_{[a}A_{b]}=2\accentset{(\{\})}{\mathfrak{D}}_{[a}A_{b]} + 2\overset{(\{\})}{w}_{[a b]}{}^{c}A_{c},
\end{equation}
where $\accentset{(\{\})}{\mathfrak{D}}_{a}A_{b}=D_{a}A_{b} - \overset{(\{\})}{w}_{a b}{}^{c} A_{c}$.  When substituting (\ref{DaCalDa}) in (\ref{stvofaction}) we see that the second term in the right-hand side of (\ref{DaCalDa}) cancels the $\tau$ term  in the first line of (\ref{stvofaction}) since $2\overset{(\{\})}{w}_{[a b]}{}^{c}=\tau_{a b}{}^{c}$ and $\delta e_{a\mu}e^{b\mu}=-e_{a\mu}\delta e^{b\mu}$. We proceed to use the product rule:
\begin{align}\nonumber
&e\overset{(4)}{\widehat{T}}{}^{a b c}2\accentset{(\{\})}{\mathfrak{D}}_{a}\delta e_{b}{}^{\mu} e_{c \mu} =\\\nonumber
&~~~=2e\accentset{(\{\})}{\mathfrak{D}}_{a}\Big(\overset{(4)}{\widehat{T}}{}^{a b c}\delta e_{b}{}^{\mu} e_{c \mu}\Big) - 2e\accentset{(\{\})}{\mathfrak{D}}_{c}\overset{(4)}{\widehat{T}}{}^{c a d}\delta e_{a}{}^{\mu} e_{d \mu}~~~ \\\label{ttofstvofaction}
&~~~~~~~ -2e\overset{(4)}{\widehat{T}}{}^{a b c}\delta e_{b}{}^{\mu} (-e_{e\mu} \Omega_{a}{}^{e}{}_{c} - \overset{(\{\})}{w}_{a c}{}^{e} e_{e \mu}).
\end{align}
The first term on the right-hand side of (\ref{ttofstvofaction}) yields a total derivative and the last line of (\ref{ttofstvofaction}) comes from expanding $\accentset{(\{\})}{\mathfrak{D}}_{a}e_{c\mu}$. Putting all together we have
\begin{align}\nonumber
&\delta_{e}\Big(-\frac{1}{2} e \overset{(4)}{\widehat{T}}{}^{a b c} \overset{(4)}{T}_{a b c} \Big) =\\\nonumber
&=\frac{1}{2}e e_{d\mu}\delta e^{d\mu} \overset{(4)}{\widehat{T}}{}^{a b c} \overset{(4)}{T}_{a b c}+ 2e \delta e_{a}{}^{\mu}e_{d \mu}\Big( \overset{(4)}{\widehat{T}}{}^{a b c} \tau^{d}{}_{b c}\\\label{variationLagW}
&~~~~~~~~~~~~~~~~~~~~~~~~~~~~ -\accentset{(\{\})}{\mathfrak{D}}_{c}\overset{(4)}{\widehat{T}}{}^{c a d}  + \overset{(4)}{\widehat{T}}{}^{b a c} \overset{(\{\})}{w}_{b c}{}^{d}\Big),
\end{align}
up to a total derivative. The $\Omega$ term of (\ref{ttofstvofaction}) combines with the $\Omega$ term of (\ref{stvofaction}) to form the $\tau$ term in the second line of (\ref{variationLagW}). The equations of motion are simply:
\begin{align}\nonumber
2 \accentset{(\{\})}{\mathfrak{D}}_{c} \overset{(4)}{\widehat{T}}{}^{c}{}_{a b}   + \,  2 \overset{(4)}{\widehat{T}}{}_{a c d} \Big( \overset{(4)}{T}{}_{b}{}^{c d} -  \overset{(4)}{K}{}^{c d}{}_{b} \Big) - \frac{1}{2} \eta_{a b} \overset{(4)}{T}_{c d e}\overset{(4)}{\widehat{T}}{}^{c d e} = 0.\\\label{eomTGflat}
\end{align}
Here we have used $\overset{(4)}{T}_{a b}{}^{c}=-\tau_{a b}{}^{c}$ and $\overset{(4)}{K}_{a b}{}^{c}=-\overset{(\{\})}{w}_{a b}{}^{c}$.
\subsection{Generic case}\label{subsecBGenericCaseEOM}
In the previous subsection we obtained the equations of motion by varying the right hand side of (\ref{FundamentalEquationtwo}). This is our goal here too. However, the variation procedure of the preceding subsection depends heavily on the form of the torsion with respect to the vielbein and if we perform that procedure for the new connections it is not easy to see how to rearrange terms in meaningful quantities. In the Weitzenb\"ock case, the equations of motion are fully described in terms of the torsion since there is no curvature. In the case of the new connections, all of them have non-zero curvature and we expect the Ricci tensor to appear in the equations of motion. Therefore, we will follow here a different route to derive the equations of motion for our generic connections, and in particular, it will yield the equations of motion for the Weitzenb\"ock case (\ref{eomTGflat}). 

We start with the following action,
\begin{equation}\label{actiongenericcase}
S=\int e R(\Gamma) dx^4,
\end{equation}
where $R(\Gamma)=g^{\mu\nu} R_{\mu \rho \nu}{}^{\rho}(\Gamma)$. We are assuming that $\Gamma$ is written in terms of first-order derivatives of the vielbein. We now vary the action with respect to the vielbein, 
\begin{align}\nonumber
&\delta S = \int \delta e R(\Gamma) dx^4\\\label{eqvarSgeneric}
 &~~~~~~~ + \int e \delta g^{\mu\nu} R_{\mu \rho \nu}{}^{\rho}(\Gamma) + \int e g^{\mu\nu} \delta  R_{\mu \rho \nu}{}^{\rho}(\Gamma).
\end{align}
The first term of (\ref{eqvarSgeneric}) is as usual,
\begin{equation}
-\int e e_{a \mu} R(\Gamma) \delta e^{a \mu}.
\end{equation}
The second term of (\ref{eqvarSgeneric}) gives:
\begin{equation}
2 \int e R_{(\mu |\rho| \nu)}{}^{\rho}(\Gamma) e^{a\nu} \delta e_{a}{}^{\mu}.
\end{equation}
The third term of (\ref{eqvarSgeneric}) can be decomposed as follows. We use the decomposition for $\Gamma$ as given by (\ref{eqDecompOfGamma}) implying a decomposition for the Riemann tensor (\ref{eqDecompOfRiemann}). Thus, we obtain: 
\begin{equation}\label{eqvarSgenericthird}
\int e g^{\mu \nu} \delta \Big(R_{\mu \rho \nu}{}^{\rho}(\{\}) + 2\overset{(\{\})}{\nabla}_{[\mu}C_{\rho]\nu}{}^{\rho} + 2C_{[\mu |\lambda|}{}^{\rho} C_{\rho]\nu}{}^{\lambda}\Big)
\end{equation}
The variation of the first term of (\ref{eqvarSgenericthird}) is obtained by varying (\ref{RiemannCurved}) with respect to $\Gamma$ with $\Gamma$ being the Levi-Civita connetion. The result is:
\begin{equation}
\int e g^{\mu\nu} \overset{(\{\})}{\nabla}_{[\mu}\delta \overset{(\{\})}{\Gamma}_{\sigma]\nu}{}^{\sigma}=0.
\end{equation}
Which yields zero, since it is a total derivative. The second and third term of (\ref{eqvarSgenericthird}) can be rewritten as follows: 
\begin{align}\nonumber
&\int  e g^{\mu\nu}\,\delta\Big( 2\overset{(\{\})}{\nabla}_{[\mu}C_{\rho]\nu}{}^{\rho}\Big) = \\\nonumber
&= \delta \Big(\int  e g^{\mu\nu}\, ( 2\overset{(\{\})}{\nabla}_{[\mu}C_{\rho]\nu}{}^{\rho})\Big) -  \int \delta e g^{\mu\nu}\, ( 2\overset{(\{\})}{\nabla}_{[\mu}C_{\rho]\nu}{}^{\rho})\\
&~~~~~~ - \int e \delta g^{\mu\nu}\, ( 2\overset{(\{\})}{\nabla}_{[\mu}C_{\rho]\nu}{}^{\rho}), 
\end{align}
and
\begin{align}\nonumber
&\int  e g^{\mu\nu}\,\delta\Big( 2C_{[\mu |\lambda|}{}^{\rho} C_{\rho]\nu}{}^{\lambda}\Big) = \\\nonumber
&= \delta \Big(\int  e g^{\mu\nu}\, ( 2C_{[\mu |\lambda|}{}^{\rho} C_{\rho]\nu}{}^{\lambda})\Big) -  \int \delta e g^{\mu\nu}\, ( 2C_{[\mu |\lambda|}{}^{\rho} C_{\rho]\nu}{}^{\lambda})\\
&~~~~~~~~ - \int e \delta g^{\mu\nu}\, ( 2C_{[\mu |\lambda|}{}^{\rho} C_{\rho]\nu}{}^{\lambda}). 
\end{align}
Putting all together
\begin{align}\nonumber
\delta S = \int e \Bigg(g_{\mu\nu}\bigg( -R(\Gamma) + 2\overset{(\{\})}{\nabla}_{[\sigma} C_{\rho]}{}^{\sigma \rho} + 2C_{[\sigma|\lambda|}{}^{\rho} C_{\rho]}{}^{\sigma\lambda}\bigg)\, +   \\\nonumber
~~~~~+ \,2 R_{(\mu |\rho| \nu)}{}^{\rho}(\Gamma) - 2\overset{(\{\})}{\nabla}_{(\mu} C_{|\rho|\nu)}{}^{\rho} + 2 \overset{(\{\})}{\nabla}_{\rho} C_{(\mu\nu)}{}^{\rho}  \\\nonumber
~~~~ -\, 2C_{(\mu|\lambda}{}^{\rho} C_{\rho|\nu)}{}^{\lambda} + 2 C_{\rho\lambda}{}^{\rho} C_{(\mu\nu)}{}^{\lambda} \Bigg)e_{a}{}^{\nu} \delta e^{a\mu} \,+  \\\label{varactiongenericcase}
~~~~~~~~~~~~~~~~~~~~~~~+\, \delta \Bigg( 2\int e C_{[\mu |\lambda|}{}^{\rho} C_{\rho]}{}^{\mu\lambda} \Bigg).
\end{align}
Since we are dealing with $\Gamma$'s such that the Ricci scalar vanishes then the action (\ref{actiongenericcase}) vanishes. This implies that $\delta S =0$ off-shell. Therefore, the the right hand side of (\ref{varactiongenericcase}) must vanish identically. On the other hand, the last term of (\ref{varactiongenericcase}) is equal to the variation of the right hand side of (\ref{FundamentalEquationtwo}) and we need to impose this variation to vanish in order to obtain the equations of motion. So, the resulting (equivalent) equations of motion are:
\begin{align}\nonumber
&\,2 R_{(\mu |\rho| \nu)}{}^{\rho}(\Gamma)  + g_{\mu\nu}\Big(2\overset{(\{\})}{\nabla}_{[\sigma} C_{\rho]}{}^{\sigma \rho} + 2C_{[\sigma|\lambda|}{}^{\rho} C_{\rho]}{}^{\sigma\lambda}\Big)\\[1pt]\nonumber
&~~~ - 2\overset{(\{\})}{\nabla}_{(\mu} C_{|\rho|\nu)}{}^{\rho} + 2 \overset{(\{\})}{\nabla}_{\rho} C_{(\mu\nu)}{}^{\rho}  -\, 2C_{(\mu|\lambda}{}^{\rho} C_{\rho|\nu)}{}^{\lambda}  \label{eomgencasecurvedindices} \\[2pt]
&~~~ + 2 C_{\rho\lambda}{}^{\rho} C_{(\mu\nu)}{}^{\lambda}\, =\, 0, 
\end{align}
where we have used $R(\Gamma)=0$.
Equivalently, in flat indices the equations of motion are:
\begin{align}\nonumber
&\,2 R_{(a |c| b)}{}^{c}(w)  + \eta_{a b}\Big(2\accentset{(\{\})}{\mathfrak{D}}_{[d} C_{c]}{}^{d c} + 2C_{[d|e|}{}^{c} C_{c]}{}^{d e}\Big) \\[1pt]\nonumber
& ~~~ - 2\accentset{(\{\})}{\mathfrak{D}}_{(a} C_{|c|b)}{}^{c} + 2 \accentset{(\{\})}{\mathfrak{D}}_{c} C_{(a b)}{}^{c}  -\, 2C_{(a|e}{}^{c} C_{c|b)}{}^{e}  \\[2pt]\label{eomgenericvarflat}
&~~~ + 2 C_{c e}{}^{c} C_{(a b)}{}^{e}\, =\, 0. 
\end{align}
Equations (\ref{eomgencasecurvedindices}) and (\ref{eomgenericvarflat}) are the equations of motion for the general case (\ref{GammaConnectioncurvedsix}) and (\ref{SpinConnectionTau}) respectively. In the case of metric-compatible connections only the contorsion tensor contributes to $C_{a b}{}^{d}$. In this case, and after some manipulations, equation (\ref{eomgenericvarflat}) gets the form:
\begin{align}\nonumber
&\,2 R_{(a |c| b)}{}^{c}(w) + 2 \accentset{(\{\})}{\mathfrak{D}}_{c} \widehat{T}{}^{c}{}_{a b}   + \,  2 \widehat{T}{}_{a c d} \Big( T_{b}{}^{c d} -  K^{c d}{}_{b} \Big)  \\\nonumber
&~~~~~~~~~~~~~ - \frac{1}{2} \eta_{a b} T_{c d e}\widehat{T}{}^{c d e} + \accentset{(\{\})}{\mathfrak{D}}_{c} T_{a b}{}^{c} + 2\accentset{(\{\})}{\mathfrak{D}}_{[a} T_{b] c}{}^{c} \\\label{eomgenericvarflatTorsion}
&~~~~~~~~~~~~~ + T_{[b}{}^{c d} T_{|c d| a]} - T_{a b}{}^{c} T_{c d}{}^{d}\,=\,0, 
\end{align}
where $\widehat{T}_{a b}{}^{c}$ was defined in (\ref{superpotential}). It is easy to see that (\ref{eomgenericvarflatTorsion}) reduces to (\ref{eomTGflat}) for the Weitzenb\"ock case,
 $\overset{(4)}{w}_{a b}{}^{c}=0$, $R_{a b c}{}^{d}(\overset{(4)}{w})=0$, $\overset{(4)}{T}_{a b}{}^{c}= - \tau_{a b}{}^{c}$, with the help of the following identity:
\begin{equation}
\accentset{(\{\})}{\mathfrak{D}}_{c} \tau_{a b}{}^{c} + 2 \accentset{(\{\})}{\mathfrak{D}}_{[a} \tau_{b] c}{}^{c} -\tau_{[b}{}^{c d} \tau_{|c d| a]}  + \tau_{a b}{}^{c} \tau_{c d}{}^{d} = 0,
\end{equation}
As a final remark, we have verified that when plugging the connections (\ref{QzeroSolOneConn})-(\ref{QzeroSolFourConn}) inside equations (\ref{eomgenericvarflatTorsion}) reproduce the same equations of motion for the vielbein as in teleparallel gravity or Einstein gravity.

\section{Summary}

\label{conclusions} \bigskip

We have analyzed conditions on a generic connection (\ref{GammaConnectioncurvedsix}) in order to reproduce equivalent formulations to Einstein Gravity. We have found in section (\ref{eqforCoeffMCcase}) that there are only four metric-compatible connections that satisfy the conditions, one of them being the Weitzenb\"ock connection. In this sense we are obtaining Teleparallel Gravity as a particular case. For a generic case that includes both torsion and non-metricity, there seems to be an infinite amount of solutions. In section (\ref{eqforCoeffNMcase}) we analyzed a particular case of a Weyl space with Weyl's vector being proporcional to $\overset{(W)}{T}_{\mu}{}^{\sigma}{}_{\sigma}$ (the vector part of the Weitzenb\"ock torsion). The connections (\ref{gammaweylspace}) have torsion and non-metricity and are related to the metric-compatible ones (\ref{QzeroSolOneConn})-(\ref{QzeroSolFourConn}) through the transformation (\ref{gaugetransfRicciScalar}).
Our generic connections (except for the Weitzenb\"ock one) present non-vanishing curvature, torsion and non-metricty so the dynamics of the theories, defined by these connections, is represented by a mixture of those quantities. This can be seen in the equations of motion (\ref{eomgencasecurvedindices}) (or (\ref{eomgenericvarflat})) where the Ricci tensor, torsion and non-metricity enters in a non-trivial way.
  
In our approach, we have the local Lorentz group broken (except for actions and equations of motion) and the interpretation we are giving is that the new connections are deformed versions of the Weitzenb\"ock connection defined over  parallelizable spacetimes.  
Of course, in this work we just analyzed the equivalence with Einstein Gravity in vaccum (no source terms coupled). So, the theories defined here, using different connections, are candidates for a complete equivalence to Einstein Gravity when sources are added. In the case of the Weitzenb\"ock connection, there is a complete equivalence with Einstein Gravity when sources are included (see \cite{deAndrade:1997gka}). We leave for future work a complete discussion of matter coupling, interpretation of the gravitational force acting on particles and geodesics.  

As a final remark, we believe our procedure here could be applied to DFT in order to find other connections determined in terms of the physical fields. However, notions of contorsion and non-metricity should be defined first in order to find a suitable decomposition as in (\ref{eqDecompOfGamma}). We also leave this issue for future work.
\section*{Acknowledgments}

The author is in debt with R. Ferraro and F. Fiorini for very useful discussions and support. We thank A. Goya for carefully reading this manuscript and suggestions and we also thank G. Aldazabal. Part of the calculations of this project have been done with Cadabra Software \cite{Peeters:2006kp}. This work was partially supported by Consejo Nacional de Investigaciones Cient\'ificas y
T\'ecnicas (CONICET).

\end{document}